\documentclass[12pt]{spieman}  % 12pt font required by SPIE;
\DeclareUnicodeCharacter{2005}{\,} % handle Unicode four-per-em space
\usepackage{graphicx}
\usepackage{textcomp}
\DeclareUnicodeCharacter{03BC}{\textmu}
\usepackage{amsmath,amsfonts,amssymb}
\usepackage{setspace}
\usepackage{tocloft}
\usepackage{comment}
\usepackage{xcolor}

\DeclareUnicodeCharacter{2212}{\textminus}
\title{Superconducting Nanowire Single-Photon Detectors for Enhanced Biomedical Imaging}

\begin{document} 
\author[a*]{Emi Cora Valmai Hughes}
\author[a,b]{Avinash Upadhya}
\author[a,c]{Kishan Dholakia}
\affil[a]{Centre of Light for Life, School of Biological Sciences, The   University of Adelaide, Adelaide, SA 5005, Australia}
\affil[b]{Robinson Research Institute, School of Biomedicine, The University of Adelaide, Adelaide, SA 5005, Australia}
\affil[c]{SUPA, School of Physics and Astronomy, University of St Andrews, St Andrews KY16 9SS, UK}

\renewcommand{\cftdotsep}{\cftnodots}
\cftpagenumbersoff{figure}
\cftpagenumbersoff{table} 

%% TO DO
% Double-check preprinted references 
% Fill placeholder citations
% Double-check with Alex about the figures

\maketitle
\begin{abstract}
\textbf{Significance:} Superconducting nanowire single-photon detectors (SNSPDs; also known as SSPDs) show enormous promise for low-light biomedical imaging by offering exceptional sensitivity, picosecond timing resolution, and broad spectral coverage.
\textbf{Aim:} This perspective evaluates the role of SNSPDs by comparing their performance with other photon-counting detectors for emerging biomedical imaging applications.
\textbf{Approach:} We outline the need for ultrasensitive detectors for biophotonics, summarize SNSPD operating principles and compare their performance with established photon-counting devices. We highlight applications in which SNSPDs enable new imaging capabilities and discuss system-level challenges and technological developments that are critical to future applications, including clinical translation.
\textbf{Results:} SNSPDs offer advantages in signal-to-noise ratio, temporal precision, and detection bandwidth, enabling deeper tissue imaging, high-precision fluorescence lifetime measurements, and quantum-enhanced imaging modalities. Advances in scalable arrays, cryogenic miniaturization, and improved signal collection are reducing barriers to widespread adoption. 
\textbf{Conclusions:} SNSPDs are poised to transform photon-limited biomedical imaging. As device performance and system integration continue to advance, their adoption in imaging platforms is expected to accelerate. Combining SNSPDs with advancements in the excitation pathway, such as structured-light excitation with Bessel beams, aberration correction, and wavefront shaping, shows promise for delivering unprecedented imaging capabilities and broadening both the preclinical and clinical utility of these detectors. 
\end{abstract}

% Include a list of up to six keywords after the abstract

% Include email contact information for the corresponding author
{\noindent \footnotesize\textbf{*}Emi Cora Valmai Hughes,  \linkable{emi.hughes@adelaide.edu.au}}

%%\begin{spacing}{2}   % use double spacing for rest of manuscript

\section{Introduction}
\label{sect:intro}
Biophotonics leverages light-matter interactions to probe biological systems. It may combine traditional optics, advanced photonics, and engineering with biology and medicine. The field aims to analyze or manipulate light-based interactions with biological molecules, living cells, tissues, and organisms. Biophotonics encompasses a multitude of applications across fields such as developmental biology \cite{ripoll_unleashing_2015, wang_label-free_2020,ilina_application_2022}, neuroscience \cite{abdelfattah_neurophotonic_2022}, oncology \cite{rowe_molecular_2022, lin_emerging_2022,balasundaram_biophotonic_2021,spyratou_advanced_2022}, cardiovascular medicine \cite{pang_diagnostic_2024,almagal_review_2025, luo_photoacoustic_2025}, and therapeutic drug monitoring \cite{tuchin_optical_2025, spyratou_biophotonic_2012}. 

Light-based imaging represents a major area of impact within biophotonics. Indeed, optical imaging $\textit{per se}$ has achieved major advances over the past few decades, providing unprecedented resolution and pushing the limits of challenging biological environments. It has surpassed the diffraction barrier \cite{klar_fluorescence_2000} and enabled resolution down to the nanometer and even Angstrom scale \cite{reinhardt_angstrom-resolution_2023} through innovations in super-resolution microscopy. In addition to offering molecular specificity and new understanding of biological processes, super-resolution microscopy now provides insights that could previously only be obtained by using electron microscopy. At a larger scale for tissue, imaging modalities such as optical coherence tomography (OCT) \cite{fujimoto_optical_2003}, light-sheet fluorescence microscopy \cite{power_guide_2017}, and multiphoton imaging \cite{konig_multiphoton_2000} are making major inroads into imaging at depth and early disease diagnosis. A common challenge across optical imaging modalities is extracting sufficient signal from the sample of interest while avoiding deleterious secondary effects such as photodamage and photobleaching. This challenge is particularly pronounced in fluorescence imaging, which exploits endogenous or exogenous fluorophores as contrast agents, especially when high signal contrast and favorable signal-to-noise ratios (SNRs) are required. Specifically, native endogenous fluorescence signals such as autofluorescence and bioluminescence \cite{yao_advances_2018} are of immense interest because they contain rich biologically relevant information, but can be challenging to capture with adequate temporal and spatial resolution \cite{peterkovic_optimizing_2025}. Extracting fluorescence signals from depth, be it from endogenous or exogenous fluorophores, remains an outstanding challenge. The recoverable fluorescence is severely reduced by the interplay of scattering and tissue absorption, and SNR is further degraded by the strong presence of background fluorescence \cite{yoon_recent_2022}. 
Conventional strategies to overcome this challenge have primarily focused on the optical excitation pathway. These strategies include multiphoton excitation \cite{konig_multiphoton_2000, helmchen_deep_2005, kobat_deep_2009}, adaptive optics \cite{rodriguez_adaptive_2021}, and complex wavefront shaping \cite{horstmeyer_guidestar-assisted_2015, yu_recent_2015}. While effective, these approaches can increase system complexity and often require sophisticated laser or light-shaping architectures for successful implementation.

A complementary paradigm is emerging that shifts the performance burden from the excitation path to the detection path. This paradigm recognizes that signals originating from deep within tissue are inherently weak and emphasizes the use of sophisticated detectors capable of photon-counting detection to capture these signals effectively. An exciting prospect is to combine this paradigm with innovations in the excitation pathway, compounding the benefits and leading to new imaging capabilities. 
A range of photon-counting detector architectures has recently emerged that can optimize the optical detection pathway in low-light imaging. Among these, superconducting nanowire single-photon detectors (SNSPDs) represent a particularly promising and intriguing prospect. SNSPDs combine unprecedented system detection efficiency \cite{chang_detecting_2021}, ultrahigh timing resolution \cite{cherednichenko_low_2021}, exceptionally low dark-count rates \cite{shibata_snspd_2017}, and broad spectral sensitivity \cite{reddy_superconducting_2020} extending deep into the infrared. Such attributes may uniquely position SNSPDs to address longstanding limitations in photon-starved modalities such as deep-tissue biomedical imaging by capturing information that conventional detectors cannot register. Consequently, SNSPDs could enable imaging modalities that were previously constrained by detector performance, including single-molecule imaging, deep-tissue fluorescence microscopy, and quantum-enhanced biophotonics. 

Although previous reviews have addressed SNSPD device architectures \cite{esmaeil_zadeh_superconducting_2021,venza_research_2025,natarajan_superconducting_2012,hu_superconducting_2015,steinhauer_progress_2021} and their applications in communication \cite{hadfield_single-photon_2009,you_superconducting_2020}, ranging \cite{li_photon_2025}, and spectroscopy \cite{lau_superconducting_2023,wollman_recent_2021}, a discussion of their biomedical applications has not been forthcoming. This paper is not intended to be a review, but rather aims to provide a perspective on this emerging topic, elucidating how SNSPDs are reshaping the landscape of deep-tissue and other low-light imaging modalities in biophotonics.

We first outline the need for single-photon detectors in biophotonics applications and then describe the fundamental operating principles of SNSPDs. We next compare the characteristics of SNSPDs with those of existing detector technologies and describe current applications in which SNSPDs enable new imaging regimes. Finally, we discuss system-level barriers that must be addressed for wider adoption. We also highlight results showing how combining innovations in the excitation pathway, such as structured light illumination and SNSPD detection can guide a future route of merit in the field.

\section{The Case for Ultra-Sensitive Detectors in Biophotonics}
The success of optical imaging for biomedical applications has been exceptional over the last few decades. Key advances include super-resolution microscopy \cite{baddeley_biological_2018} and novel imaging geometries to achieve high-resolution volumetric capabilities, such as light-sheet fluorescence microscopy \cite{stelzer_light_2021}. A frontier in the field that remains highly challenging is defined by photon scarcity. For example, recovering signals from weak endogenous autofluorescence \cite{monici_cell_2005} can be extremely challenging. A prime topic is low-light imaging at depth within highly scattering or absorptive tissues \cite{yoon_recent_2022}. As imaging depth increases, challenges arise in ensuring that light is focused deep within tissue without aberration. An associated challenge is retrieving emitted signals from depth, as these photons also undergo scattering. 

Historically, progress in imaging at depth has largely centered on the excitation pathway. Multiphoton microscopy, for example, has yielded spectacular demonstrations where scattering is dramatically reduced by moving to longer wavelengths. This shift has resulted in a wealth of important data. For instance, two-photon imaging has become widely used in intact whole mouse brains \cite{ahrens_whole-brain_2013}, demonstrating its ability to penetrate deeply into living tissue while preserving subcellular resolution. One can extend this concept to three-photon microscopy for even greater depth penetration \cite{cheng_comparing_2020}. Notably, moving to either two-photon or three-photon excitation requires the use of expensive, pulsed light sources that may also have large physical footprints \cite{xu_multiphoton_2024}. Their optimal \textit{in vivo} performance requires careful optimization of laser pulse energy, repetition rate, and average power. Although low-repetition-rate, high-peak-power lasers can reduce cumulative heating \cite{gasparoli_is_2020}, consideration must also be taken to limit the risks of nonlinear tissue damage \cite{resan_two-photon_2014}, fluorophore saturation \cite{yildirim_functional_2019}, and physiological perturbation \cite{bakker_intravital_2022,thornton_long-term_2024}. 

Fiber-based delivery provides another means of illuminating and recovering signals from deep within tissue. However, the physical footprint of the fiber can cause tissue damage. Such trauma may be minimized through the use of ultrathin multimode fibers (MMFs) \cite{wen_single_2023}. These fibers normally produce a speckle output; however, with knowledge of the transmission characteristics of the fiber, complex wavefront shaping can convert the typically random output into a deterministic and useful illumination profile. For example, this approach enables point scanning \cite{cao_controlling_2023} and the use of structured light fields \cite{mouthaan_generating_2025}, providing minimally invasive access for \textit{in situ} imaging. An exciting example is a 110~$\mu$m-diameter MMF capable of imaging the entire mouse brain, that is, through more than 5 mm of tissue
\cite{stiburek_110_2023}. However, MMFs remain sensitive to bending and environmental conditions, and their spatial resolution is fundamentally constrained by the achievable numerical aperture (NA) of the fiber \cite{wen_single_2023}.

While these methods have substantially advanced deep-tissue imaging, it is also worthwhile to explore approaches that use advanced photodetectors to maximize data collection and processing in photon-starved studies. It is this area we revert to here. SNSPDs are a prime example of an emerging detection platform capable of addressing the challenges of high-performance imaging in photon-scarce scenarios. They are uniquely positioned to overcome the bottlenecks imposed by photon scarcity. SNSPDs not only complement existing excitation methods but also enable new regimes of imaging and spectroscopy that remain inaccessible with conventional detectors. To understand how SNSPDs achieve this, we next describe their operational mechanisms and key performance metrics and benchmark their performance against conventional photon-counting detectors.

\section{Fundamental Working Principles}
\label{sect:fundamental}
\subsection{Operational Mechanism}
\begin{figure}
    \centering
    \includegraphics[width=0.8\linewidth]{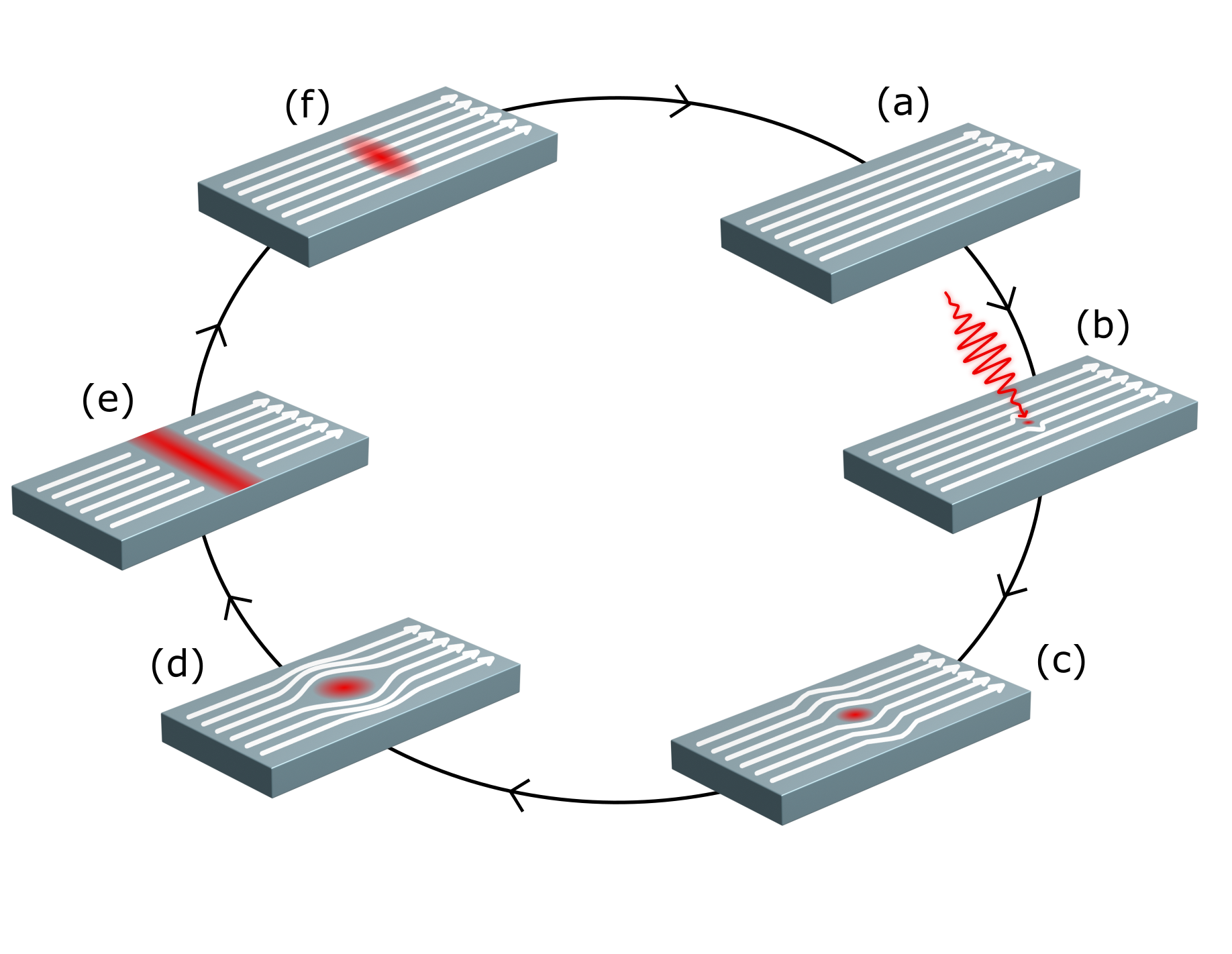}
    \caption{Working principle of an SNSPD at a macroscopic level, looking at a cross-section of the nanowire. (a) The nanowire is stabilized below the critical temperature and biased with a current (depicted as white arrows), keeping it in a superconducting state. (b) A photon is absorbed, creating a small resistive hotspot. (c) The current is diverted around the hotspot, flowing along the outer edge of the nanowire. (d) As the current diversion continues, the local current density surrounding the hotspot increases until it exceeds the critical current density required for superconductivity. (e) This results in the formation of a resistive barrier across the entire width of the nanowire. The resistance rises rapidly, resulting in a measurable voltage pulse. The current flow is blocked, and the external circuit is used to shunt the bias current. (f) The reduced current allows the resistive region to cool and collapse, returning the nanowire to a fully superconducting state. This diagram was inspired by Natarajan et al. \cite{natarajan_superconducting_2012} and Lau et al. \cite{lau_superconducting_2023} respectively.}
    \label{fig:working_principles}
\end{figure}

SNSPDs detect single photons through the principle of current-biased superconductivity disruption. The detector consists of an ultrathin niobium nitride (NbN) nanowire, typically 100~nm wide and 5-10~nm thick, stabilized well below its superconducting critical temperature and biased with a direct current just below its critical current (Fig.~\ref{fig:working_principles}(a)). The nanowire is arranged in a meander geometry to maximize the filling factor of the active area and to improve the coupling efficiency \cite{xiong_reducing_2022}. Upon photon absorption by the nanowire  (Fig.~\ref{fig:working_principles}(b)), sufficient photon energy is deposited locally to break hundreds of Cooper pairs, creating a localized resistive hotspot  (Fig.~\ref{fig:working_principles}(c)). The hotspot forces the current to divert around the affected region (Fig.~\ref{fig:working_principles}(d)), thereby increasing the local current density. As the hotspot expands to span the full width of the nanowire, the current may exceed the switching threshold, triggering the formation of a resistive section across the wire (Fig.~\ref{fig:working_principles}(e)). The resistance rises rapidly from zero (the superconducting state) to several k$\Omega$ on a picosecond timescale, producing a measurable voltage pulse across the nanowire. After the current is diverted, the hotspot cools (Fig.~\ref{fig:working_principles}(e)) and the nanowire returns to its superconducting state, typically within 1~ns (Fig.~\ref{fig:working_principles}(a)) \cite{you_superconducting_2020}. The detector is then ready to register another photon. 

\subsection{Key Performance Metrics and Device Characteristics}
When evaluating the performance of conventional cameras or detectors, parameters like the pixel fill factor and pixel cross-talk are commonly considered. However, the traditional metrics used for cameras are not appropriate for single-pixel detectors. We now consider metrics relevant to the operation of photon-counting detectors. 

\textbf{System detection efficiency (SDE):}
The probability that a photon is detected and registered as a count. This metric is critical in photon-counting applications, as it directly influences SNR, acquisition time, and image fidelity. In an idealized form, SDE is expressed as $\eta = \frac{R_{detected}}{R_{incident}}$, where $R_{incident}$ is the rate of photons incident on the detector and $R_{detected}$ is the rate of registered detection events. In practice, SDE is reduced by various losses and is expressed as the product of three components: 
$\eta = \eta_{coupling}\times\eta_{absorption}\times\eta_{internal}$. The coupling efficiency, $\eta_{coupling}$, is the probability that a photon reaches the detector. It is influenced by geometric factors such as the acceptance angle of the collection optics and any mismatch between the coupling fiber diameter and the detector area.  The absorption efficiency, $\eta_{absorption}$, represents the probability that a photon reaching the detector's active area is absorbed. Following absorption, the internal quantum efficiency, $\eta_{internal}$, denotes the probability that the absorbed photon triggers a measurable electrical signal. For SNSPDs, $\eta_{internal}$ is often inferred indirectly, as isolating it from the overall system loss remains experimentally challenging \cite{reddy_superconducting_2020}.

\textbf{Dark count rate (DCR):}
The number of spurious detection events per unit time in the absence of incident photons. These events arise from both intrinsic mechanisms (e.g., thermal excitation in the detector) and extrinsic sources such as blackbody emission, stray-light leakage, or thermal radiation from the coupling optics \cite{raj_waveguide_2025}. Minimizing the DCR is essential for accurate operation in low-light applications such as fluorescence correlation spectroscopy and single-molecule imaging. 

\textbf{Timing jitter:}
The statistical variation in the delay between photon absorption and the generation of the detector's output electrical pulse \cite{hadfield_single-photon_2009}. It is typically represented by a probability density function and determines the precision of photon arrival time measurements. High jitter degrades the temporal resolution of time-resolved techniques such as fluorescence lifetime imaging microscopy (FLIM), hindering the ability to distinguish fluorophores with similar lifetimes \cite{hirvonen_fast_2020}.

\textbf{Spectral bandwidth:}
This describes the wavelength range over which a detector is sensitive, primarily determined by its constituent materials and optical design \cite{hadfield_single-photon_2009}. Matching the detector's bandwidth to the emission spectrum of the signal source is critical for achieving high detection efficiency for any biophotonics application. \newline

%\textbf{Dynamic range}
In addition to performance metrics, intrinsic device characteristics should also be considered. 
\newline

\textbf{Afterpulsing:}
This refers to the generation of spurious electrical pulses that follow a true photon detection event. It results from trapped charge carriers that are released after the detector is re-biased. This phenomenon is prevalent in avalanche-based detectors such as single-photon avalanche diodes (SPADs), where carriers trapped during the avalanche process are thermally released and then re-trigger the circuit, mimicking a new detection event \cite{wang_advances_2025}. Therefore, afterpulsing results in an overestimation of the total count rate. To mitigate afterpulsing, these detectors impose a dead time for carrier dissipation. However, longer dead times reduce the detector's duty cycle and limit the achievable count rate \cite{ziarkash_comparative_2018}. In contrast to SPADs and photomultiplier tubes (PMTs), SNSPDs are inherently free from afterpulsing \cite{yokota_low-light_2021}. 
 %After detecting a photon, SNSPDs return to the superconducting state without being retriggered, making them highly advantageous for precise time-resolved fluorescence applications \cite{yokota_low-light_2021}.

\textbf{Latching:} 
This phenomenon occurs when an SNSPD remains trapped in a resistive state after photon absorption, meaning that further measurements cannot be taken. Such latching can occur due to electrical noise, thermal fluctuations, and, most importantly, a high influx of photons within a short time interval \cite{annunziata_reset_2010}. This makes SNSPDs highly sensitive to environmental noise; for example, stray light can induce latching even when the detected count rate is well below the maximum count rate (MCR) \cite{liu_nonlatching_2012}. Latching also has implications for the dynamic range of the SNSPD, which is discussed in section \ref{sect:challenges} of this perspective. 

\subsection{Comparison of the Advantages and Disadvantages Among Single-Photon Detectors}
In this section, we compare the performance of three classes of popular, highly sensitive detectors used in photon-counting experiments: photomultiplier tube (PMTs), SPADs, and SNSPDs. Although imaging sensors such as scientific complementary metal-oxide-semiconductor (sCMOS) and electron multiplying charge-coupled device (EMCCD) cameras have greatly advanced low-light imaging and have been used for single-photon experiments \cite{peterkovic_optimizing_2025}, they are limited by readout noise and photon-number resolution and are therefore not considered here.

\textbf{Photo multiplier tube (PMT)}
A PMT consists of a photocathode and a series of dynodes enclosed in a vacuum tube. When a photon strikes the photocathode, it releases a photoelectron that is multiplied through successive dynodes to produce a measurable current at the anode. PMTs offer a large active area, fast timing response, and low DCRs. However, the quantum efficiency is typically limited to 10-40\% \cite{eisaman_invited_2011}, and the spectral sensitivity is largely restricted to wavelengths below 800~nm \cite{lawrence_enhanced_2008}. PMTs are fragile, and their sensitivity to magnetic fields can complicate integration into compact or multimodal instruments.

\textbf{Single-photon Avalanche Diode (SPAD)}
SPADs operate in Geiger mode, in which the absorption of a single photon triggers an avalanche breakdown in a reverse-biased p-i-n junction. They offer compact form factors, high SDE (up to ~70\% in the visible range), and picosecond-level timing resolution (30-100~ps) \cite{bruschini_single-photon_2019}, making them ideal for time-correlated single-photon counting (TCSPC) applications. However, SPADs exhibit moderate DCRs, as well as afterpulsing, which leads to long dead times (typically tens to thousands of nanoseconds) that limit their overall SDE \cite{albeck_dead-time_2025}. Indium gallium arsenide (InGaAs) SPADs extend detection to telecom wavelengths (up to 1.55~$\mu$m), but they require active cooling to suppress thermal noise \cite{meng_ingaasinalas_2016}. Continued advances aim to reduce noise and dead time and improve near-infrared (NIR) performance, though challenges in array readout speed persist \cite{ceccarelli_recent_2021}.

The suitability of a single-photon detector for a biophotonics application is determined by a balance between performance, practicality, and scalability. As summarized in Table~\ref{tab:detector_comparison}, PMTs, SPADs, and SNSPDs each present distinct trade-offs. PMTs offer robust, low-noise detection; SPADs offer compact and increasingly sophisticated array-based architectures; and SNSPDs offer superior SDEs and timing resolution at the cost of a large size, weight, power consumption, and cost (SWaP-C), as well as the need for cryogenic operation. These contrasting characteristics shape the practical feasibility of different detectors for each single-photon modality. Although most commercial biophotonic imaging systems operating at low light levels rely on PMTs, the development of SPADs has been instrumental in driving progress across a range of advanced imaging and spectroscopy methods \cite{bruschini_single-photon_2019}, often replacing PMTs in these applications \cite{chunnilall_metrology_2014}. SNSPDs are expected to further advance the state of the art in sensitivity, timing precision, and spectral range.
For a more in-depth analysis, the reader is referred to existing reviews covering SNSPDs \cite{esmaeil_zadeh_superconducting_2021}, SPADs \cite{bruschini_single-photon_2019}, and single-photon detectors in general \cite{hadfield_single-photon_2009,chunnilall_metrology_2014, wang_advances_2025,eisaman_invited_2011}. 

\begin{table}[hbt!]
\begin{center}       
\begin{tabular}{|l|l|l|}
\hline
\rule[-1ex]{0pt}{3.5ex}  \textbf{Detector type} &  \textbf{Advantages} & \textbf{Disadvantages} \\
\hline
\rule[-1ex]{0pt}{3.5ex}  \textbf{PMT} & 
High gain and low electronic noise & Sensitivity to magnetic fields \\
& Fast temporal response & High operating voltage  requirements\\
& Large active area & Vacuum-tube architecture\\
%(limits their lifetime reliability and scalability)\cite{eisaman_invited_2011} 
&& Easily damaged by excessive light exposure \cite{giacomelli_evaluation_2019}\\
\hline 
\rule[-1ex]{0pt}{3.5ex}  \textbf{SPAD} &
Availability of megapixel arrays &  Dead time typically 10~ns to 10~$\mu$s \cite{eisaman_invited_2011}\\
& High dynamic range & Afterpulsing requires gating or quenching \\
& Integration with sCMOS technology \cite{eisaman_invited_2011} & \\
& Sub-nanosecond timing resolution & \\
& Operation at room temperature & \\
\hline 
\rule[-1ex]{0pt}{3.5ex}  \textbf{SNSPD} & Near-unity system detection efficiency & Large size, weight, power, and cost (SWaP-C) \\
& Ultra-low dark counts & Small active area  \\
& Excellent timing resolution & Requires cryogenic cooling \\
& No afterpulsing and negligible dead time & Polarization sensitivity \\
& Broad spectral range up to 10~$\mu$m \cite{eisaman_invited_2011} & Limited availability of pixelated arrays  \\
& & Complex fabrication process \\
& & Latching \\
\hline 
\end{tabular}
\end{center}
\caption{Comparison of the advantages and disadvantages of a PMT, SPAD, and SNSPD in the context of biophotonics applications. As the field continues to evolve, the parameters presented reflect the current state of the technology and are included for indicative purposes. For some of the latest performance parameters, references are presented.}
\label{tab:detector_comparison}
\end{table} 

\section{Current Opportunities Enabled by SNSPD Technology}
\label{sect:opportunities}
We now highlight several imaging opportunities in biophotonics where SNSPDs provide distinctive advantages over alternative detectors. These include imaging in the NIR window for deeper tissue penetration and reduced tissue autofluorescence, measuring ultrafast photon-arrival timing to probe molecular dynamics, and low-noise detection for correlation spectroscopy and quantum imaging. 

\subsection{Short-Wave Infrared Fluorescence Microscopy and Large Field-of-View Microscopy for Imaging Tissue at Depth} 
A fundamental limitation in optical microscopy is its shallow penetration depth, which is largely constrained by light scattering and autofluorescence \cite{wang_vivo_2024}. Short-wave infrared (SWIR) fluorescence imaging, particularly emission in the near-infrared-II (NIR-II, $\sim$1.0-1.7~$\mu$m) spectral window, offers photophysical properties that are optimal for biological imaging. These include deeper tissue penetration through reduced scattering \cite{welsher_route_2009, smith_second_2009, bashkatov_optical_2005}, negligible tissue autofluorescence \cite{del_rosal_strategies_2018}, and improved superficial contrast \cite{carr_absorption_2018}. Longer-wavelength excitation also lowers photon energy, thereby reducing phototoxicity and broadening the safety margin for \textit{in vivo} imaging \cite{park_golden_2017, upputuri_photoacoustic_2019}. 

Despite these advantages, the promise of SWIR imaging has long been constrained by detector technology. Conventional detectors such as InGaAs photodiodes, SPADs, and PMTs suffer from low sensitivity and high noise in the SWIR region. SNSPDs overcome these barriers by combining high SDE, ultra-low DCR, and extended sensitivity beyond 2~$\mu$m, establishing them as an enabling detector for low-light, high-fidelity deep-tissue imaging.

Pioneering work by Xia et al. \cite{xia_short-wave_2021} reported the first demonstration of SWIR confocal fluorescence imaging deep in mouse brain enabled by SNSPDs. Using a continuous-wave (CW) 1.31~$\mu$m excitation source and a custom-built SNSPD optimized for the 1.7~$\mu$m window, they achieved one-photon confocal imaging approaching 1.7~mm below the cortical surface. The customized SNSPD exhibited a detection efficiency more than 50 times higher than that of an InGaAs PMT at wavelengths beyond 1.6~$\mu$m, establishing one-photon SNSPD-enabled SWIR microscopy as a viable route for deep-tissue imaging.

\begin{figure}
    \centering
    \includegraphics[width=0.7\linewidth]{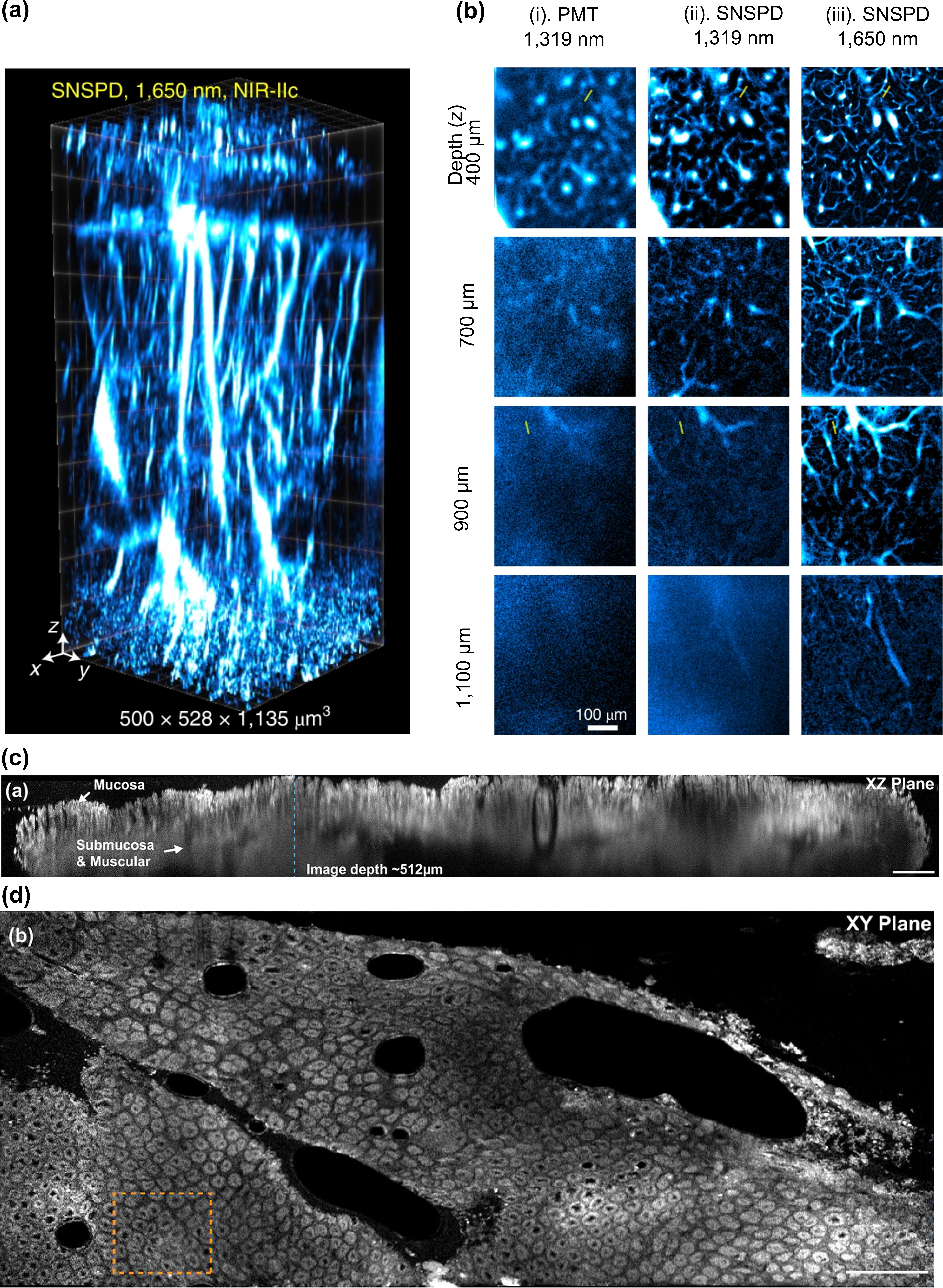}
    \caption{SWIR confocal imaging using an SNSPD. (a) Volumetric images of blood vessels in an intact mouse head (from scalp to cortex), acquired with 5~$\mu\text{m}$ axial scan increments. A 1.65~$\mu$m laser was used for excitation, and fluorescence was collected in the 1.8-2~$\mu$m window. Imaging was performed 30~minutes after intravenous injection of quantum dots. (b) High-resolution confocal images at various depths using PMT and SNSPD detectors. (i) PMT with 1.32~$\mu$m excitation and 1.5-1.7~$\mu$m collection. (ii) SNSPD with the same excitation and collection wavelengths as the previous panel. (iii) SNSPD with 1.65~$\mu$m excitation and 1.8-2~$\mu$m collection. All lasers operated at 28.5~mW at the mouse head surface. Results are representative of three mice (BALB/c, female, 3 weeks old). (c)  Stitched xz scan (4,500~$\mu \text{m}$ $ \times$ 512~$\mu \text{m}$) of mouse colon tissue using SNSPD with 1 mW excitation. The 512~$\mu \text{m}$  imaging depth spans the mucosa, submucosa, and muscular layers. Data are composed of 16 individual xy-plane scans. (d) Stitched xy scan (2.3~mm $\times$ 0.95~mm) of colon tissue at a depth of approximately 100~$\mu \text{m}$. Credit: Panels a-b adapted with permission from \cite{wang_vivo_2022} © Springer Nature Ltd.; panels c-d adapted with permission from \cite{liu_superconducting_2024} © Optica Publishing Group.}
    \label{fig:SWIR_Confocal}
\end{figure}

Longer-wavelength excitation and fluorescence were demonstrated by Wang et al. \cite{wang_vivo_2022} using a multiwavelength SNSPD for detection of fluorescence emission up to 2~$\mu$m. This was combined with PbS/CdS quantum dots emitting at 1.88~$\mu$m as fluorescent labels. Using one-photon excitation at 1.65~$\mu$m, they achieved an imaging depth of $\sim$1.1~mm through an intact, non-cleared mouse head. 
Figure \ref{fig:SWIR_Confocal}(a) demonstrates the volumetric capabilities of the imaging system, capturing detailed three-dimensional vascular structures. Figure \ref{fig:SWIR_Confocal}(b) compares confocal images acquired using a conventional PMT (i) with an SNSPD (ii) under identical 1.31~$\mu$m excitation. The SNSPD provides markedly higher contrast and improved structural clarity at greater depths. Furthermore, by extending the excitation to 1.65~$\mu$m (iii) reveals additional fine structures that are not visible at 1.31~$\mu$m, while also improving the overall image quality throughout the imaged volume. This study demonstrated the benefits of the linear depth-dependent signal decay of one-photon excitation, in contrast to the quadratic or cubic dependencies in two-photon and three-photon microscopy, respectively. The study was the first to achieve both excitation and emission beyond 1.5~$\mu$m, demonstrating that NIR-II fluorophores paired with SNSPDs enable high-contrast, noninvasive \textit{in vivo} imaging deep into biological tissue.

One of the key challenges in photon-counting device design is scaling from a single-pixel detector to an array of detectors comprising numerous pixels. Array detectors extend beyond single-pixel detection and, in a biophotonics context, can offer several advantages: they provide spatial resolution, enhance the dynamic range by parallelizing the intensity measurement \cite{hao_compact_2024}, and increase the effective detector area. 

Application of an SNSPD array was demonstrated by Tamimi et al. \cite{tamimi_deep_2024} for high-throughput multiphoton imaging, establishing the feasibility of parallel photon counting using pixelated SNSPDs, which is a step toward scalable, high-dynamic-range imaging. Two-photon excitation in deep tissue commonly uses low-repetition-rate femtosecond pulsed lasers, operating in a single-pulse-per-pixel regime \cite{tamimi_deep_2024}. This is significantly aided by large-area array detectors, which provide the dynamic range needed to visualize varying fluorescence levels across pixels and improve detection efficiency through the additional collection of multiply scattered fluorescence photons. The study developed a 6$\times$6 SNSPD array, yielding a total active area of $\sim$3,600~$\mu$m$^2$. This design improved the light collection efficiency of scattered fluorescence photons and increased dynamic range by spatially distributing the detected signal across multiple pixels. Using an NIR organic dye with two-photon excitation at $\sim$1.7~$\mu$m and emission at $\sim$1.1~$\mu$m, the system achieved \textit{in vivo} imaging of mouse brain vasculature to depths of $\sim$1.1~mm. This constitutes the first biological demonstration of multiphoton imaging using an integrated SNSPD array, marking a key step toward scalable SNSPD-based deep-tissue imaging platforms.

Clear steps toward the clinical translation of a single-pixel SNSPD-enabled confocal microscope have also been demonstrated. NIR-II imaging was demonstrated in a clinical setting by Liu et al. \cite{liu_superconducting_2024} through the development of a portable confocal microscope incorporating a single-pixel SNSPD device. Using indocyanine green (ICG) as a fluorophore, their system achieved imaging depths in tissue up to 512~$\mu$m (Fig.~\ref{fig:SWIR_Confocal}(c,d)). The SNSPD's high sensitivity enables excitation powers as low as 1~mW, thereby avoiding the risk of tissue damage and demonstrating the clinical promise of SNSPD-based surgical guidance.

\begin{figure}
    \centering
    \includegraphics[width=1\linewidth]{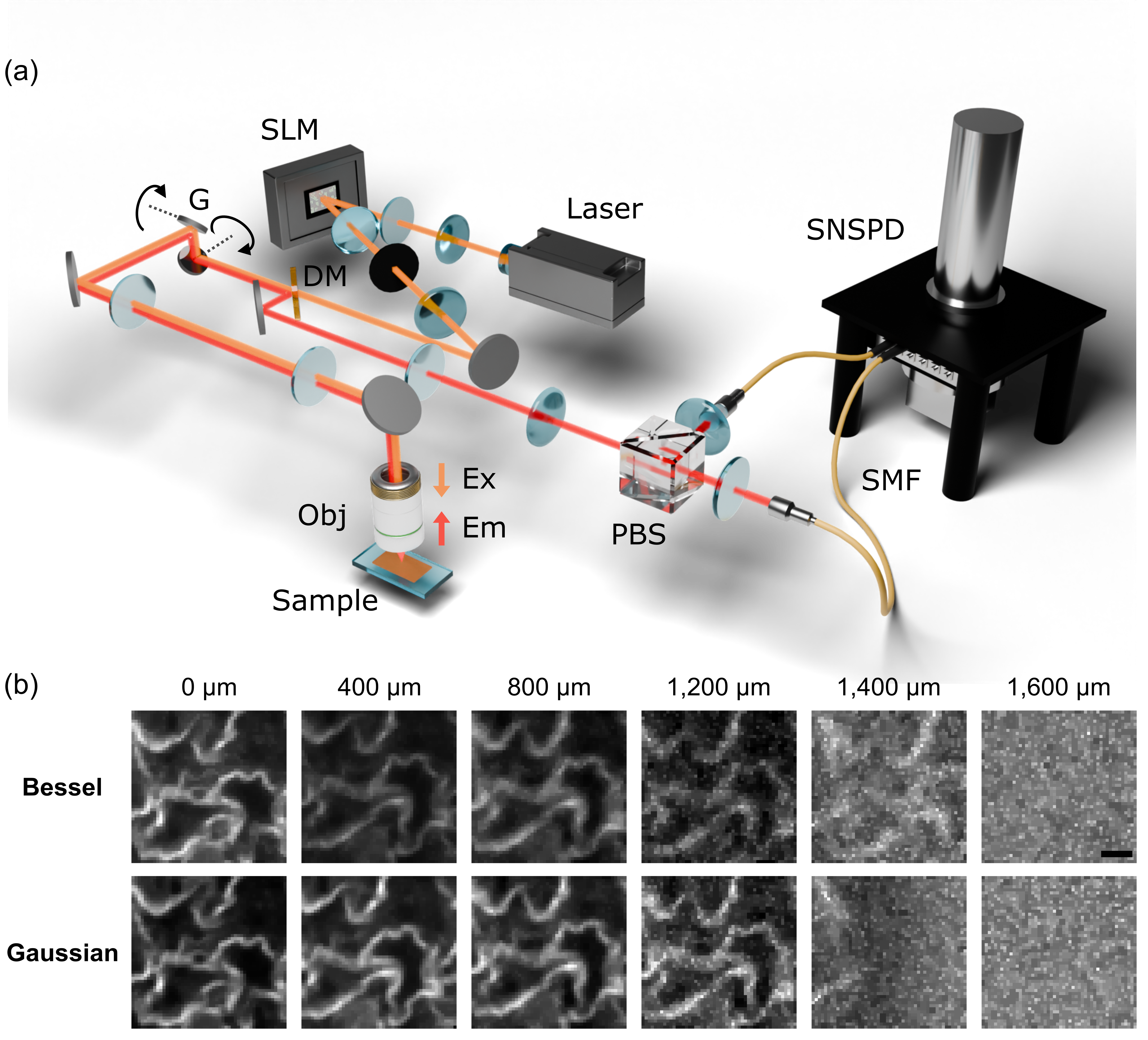}
    \caption{(a) Schematic of the near-infrared (NIR) confocal microscope equipped with wavefront shaping and superconducting nanowire single-photon detector readout. The excitation path consists of a 1.31~$\mu$m continuous-wave laser, beam expander, spatial light modulator (SLM), dichroic mirror (DM), xy galvanometric scanner (G), and objective lens (Obj), which focuses light onto the sample. The collection path comprises the objective (Obj), x-y galvanometric scanner (G), dichroic mirror (DM), polarizing beam splitter (PBS), achromatic doublet, single-mode fiber (SMF), and SNSPD. The excitation path is represented in orange and the emission path is represented in red. (b) Depth-resolved confocal fluorescence images of a monolayer of NIR quantum dots overlaid with successive layers of scattering phantom tissue added in 400~$\mu\text{m}$ increments. Images were acquired under Bessel beam and Gaussian beam illumination. Scale bar 40~$\mu$m}
    \label{fig:OpticalSchematic}
\end{figure}

To signpost future directions, in particular, to highlight the combination of excitation and collection engineering, we conducted a pilot experiment comparing conventional Gaussian beams with structured light. Specifically, we investigated Bessel beam illumination in one-photon confocal microscopy. The Bessel beam has interesting properties such as propagation invariance \cite{bouchal_self-reconstruction_1998, vasnetsov_self-reconstruction_2000, mcgloin_bessel_2005, garces-chavez_reconstruction_2004}, representing one of many avenues for potential improvements in microscopy at depth. We imaged a monolayer of NIR-II quantum dots (Nirmidas, NBDY-0018B), embedded beneath agarose-based scattering phantoms to mimic brain tissue \cite{hong_through-skull_2014}(scattering coefficient was measured to be $\mu_s = 2.57 ~m{m}^{-1}$). Figure~\ref{fig:OpticalSchematic}(a) shows the optical layout. A 1.31~$\mu$m CW laser overfills a spatial light modulator (SLM, Meadowlark), and the first diffracted order is relayed through 4f beam-compression lenses and spatially filtered with an iris. Here, the SLM is used to project a Bessel or Gaussian beam onto the sample plane. An xy galvanometric scanner provides point scanning, and the beam is focused onto the sample using a water-immersion objective (Olympus XLPLN25XWMP2). Fluorescence emitted by the quantum dots is collected in a similar geometry to that used in the study by Xia et al. \cite{xia_short-wave_2021}. As described in Section \ref{sect:challenges}, this differential detection scheme suppresses noise sources common to both channels while maximizing the signal collection across the two polarizations. 

The total power in the central lobe of the Gaussian ($2\omega_0$) and Bessel ($2r_0$) beams is matched to ensure equal illumination power at the sample plane. Using a Bessel beam, we observed imaging performance comparable to that obtained under Gaussian illumination. This work represents the first application of structured light in the SWIR regime using an SNSPD and highlights its enabling role in low-light biological imaging scenarios. In the future, not only Bessel beams but also strategies such as adaptive optics and wavefront shaping could be of use in this area. The high gain, low noise, and excellent timing resolution of SNSPDs offer a compelling platform for further exploration of such illumination strategies in this spectral window.

\begin{figure}[hbt!]
    \centering
    \includegraphics[width=0.8\linewidth]{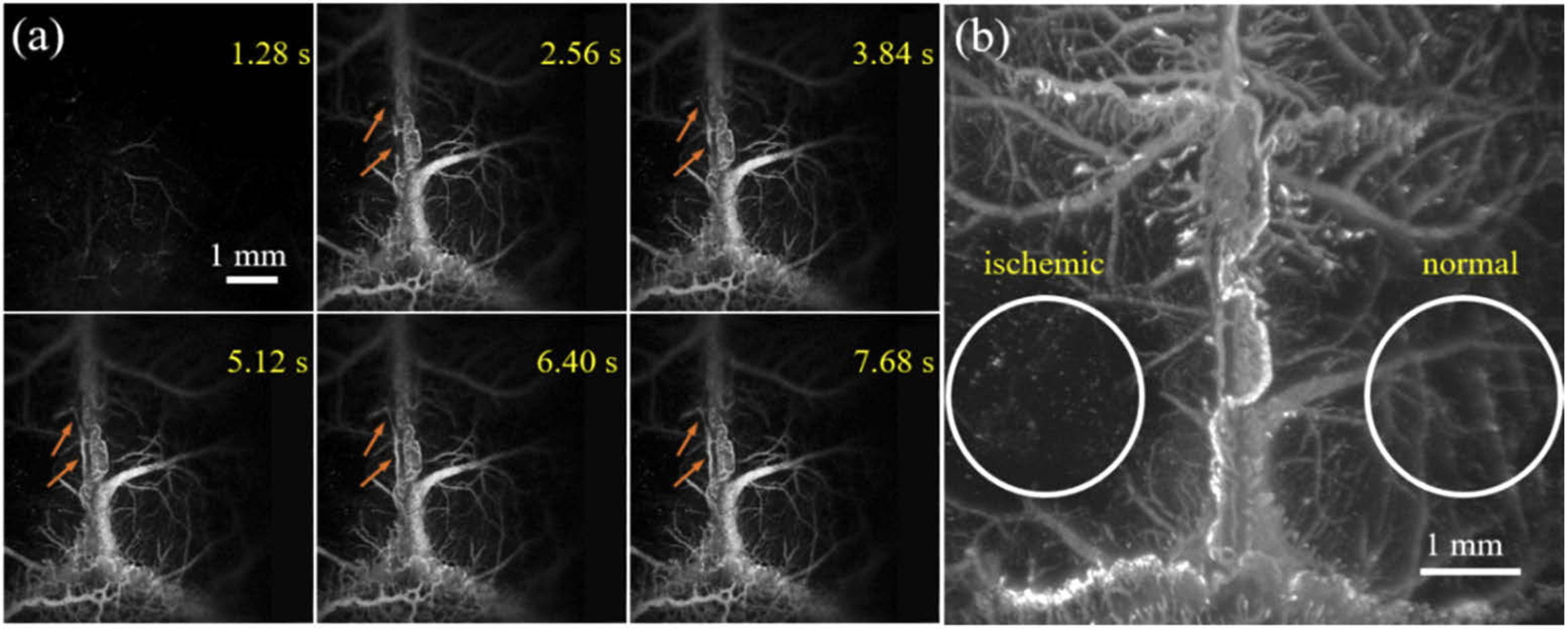}
    \caption{In vivo through-skull vasculature imaging of a cerebral ischemia mouse, where thrombosis was photochemically induced. (a) In vivo time-lapse imaging. Scale bar 1~mm. (b) The maximum intensity projection image of the cerebral vasculature from depths of 0~mm to 2.5~mm. Scale bar 1~mm. Panels a-b adapted with permission from Liao et al.\cite{liao_depth-resolved_2020} © Optica Publishing Group.}
    \label{fig:opticsExpressMesoscope}
\end{figure}

SNSPDs have also been applied to large field-of-view (FOV) imaging. In the system reported by Liao et al. \cite{liao_depth-resolved_2020}, a FOV of 7.5$\times$7.5~mm with a lateral resolution of 6.3~$\mu$m was achieved, enabling \textit{in vivo} cerebrovascular imaging of a mouse. As shown in Fig.~\ref{fig:opticsExpressMesoscope}, the system achieved time-lapse imaging over a large FOV, with imaging speed limited by the mechanical scanning of the excitation beam rather than detector performance.

Together, these examples underscore the unique ability of SNSPDs to overcome traditional performance trade-offs in SWIR imaging. As multipixel, broadband, and large-active-area SNSPDs mature and integration with imaging systems improves, we anticipate a new class of biophotonic tools capable of resolving biological processes at depth and with quantum-limited sensitivity beyond that of standard epifluorescence imaging.

\subsection{Single-Photon Spectroscopy}
High-performance single-photon spectrometers are increasingly indispensable for probing photosensitive biochemical processes. SNSPDs have enabled advancements across multiple spectroscopy modalities. They have been integrated into diffuse correlation spectroscopy (DCS) \cite{colombo_-vivo_2021,damagatla_interstitial_2023,parfentyeva_fast_2023,ozana_superconducting_2021,poon_first--clinical_2022,sieno_oxyhemoglobin_2025}, Raman spectroscopy \cite{sidorova_fiber-dispersive_2021,toussaint_proof_2015}, fluorescence correlation spectroscopy \cite{yamashita_fluorescence_2014}, and on-chip reconstructive spectroscopy \cite{zheng_-chip_2023}, illustrating their versatility in both fundamental and applied biophotonics. Furthermore, advanced spectrometers with improved quantum efficiency can be used to enhance other imaging modalities, such as Fourier-domain OCT \cite{boer_twenty-five_2017} and hyperspectral imaging \cite{yoon_hyperspectral_2022}. %

Among these applications, DCS has attracted particular attention because of its ability to noninvasively assess deep-tissue perfusion from temporal fluctuations in diffusely scattered photons. These fluctuations arise from changes in the interference pattern at the detector caused by moving tissue scatterers. Given this relevance, we focus specifically on DCS. 

Time-domain DCS (TD-DCS) with an SNSPD was demonstrated by V. Parfentyeva et al. \cite{parfentyeva_fast_2023} for \textit{in vivo} measurements to obtain a gated pulsatile blood-flow index on the adult human forehead. Compared to a commercial SPAD detector, the SNSPD demonstrated an improved timing resolution, a sixfold increase in count rate, no afterpulsing, and negligible background noise. Furthermore, the SNSPD achieved higher sensitivity between 1.05 and 1.1~$\mu$m, which corresponds to local minima in water absorption that are known to improve the SNR \cite{carp_diffuse_2023}.

Despite their large size, weight, power, and cost (SWaP-C), SNSPDs have shown the capability to enable clinical, state-of-the-art time-domain DCS. Although maximum permissible exposure is limited in DCS, the increased detection efficiency enabled by SNSPDs will advance TD-DCS for \textit{in vivo} applications. Recent studies have extended the use of SNSPDs to neurointensive care settings, enabling continuous pulsatile blood-flow monitoring in a traumatic brain injury patient \cite{poon_first--clinical_2022}. Taken together, these results highlight the trajectory of SNSPD-based spectroscopy toward real-world clinical applications.

\subsection{Time-Resolved Fluorescence Lifetime Measurements}
Unlike conventional fluorescence microscopy, which measures the absolute intensity emitted by fluorescent molecules, FLIM measures the time that a fluorophore is in the excited state before it decays to the ground state \cite{wallrabe_imaging_2005}. In addition to spatially localized fluorescence intensity, the fluorescence lifetime can provide environmental information, such as viscosity, temperature, pH, oxygen concentration, or molecular interactions. FLIM measurements of native fluorescent compounds within cells can also reveal biologically relevant information, such as the metabolic state of cells \cite{chow_viewing_2024}. 

The timing resolution and sensitivity of the detector are particularly important for FLIM, as they determine the range of fluorescence decay times that can be measured and hence the ability to distinguish between different fluorophores. FLIM is a self-referenced measurement because the fluorescence lifetime is extracted from the temporal decay profile of the emitted fluorescence, rather than the absolute intensity. Consequently, FLIM does not require the calibration steps typically necessary for intensity-based experiments \cite{datta_fluorescence_2020}. A range of detectors has been applied to FLIM, from point detectors such as PMTs and SPADs to widefield photodetectors such as EMCCDs and sCMOS cameras \cite{yokota_low-light_2021}. 

\begin{figure}[hbt!]
    \centering
    \includegraphics[width=1.0\linewidth]{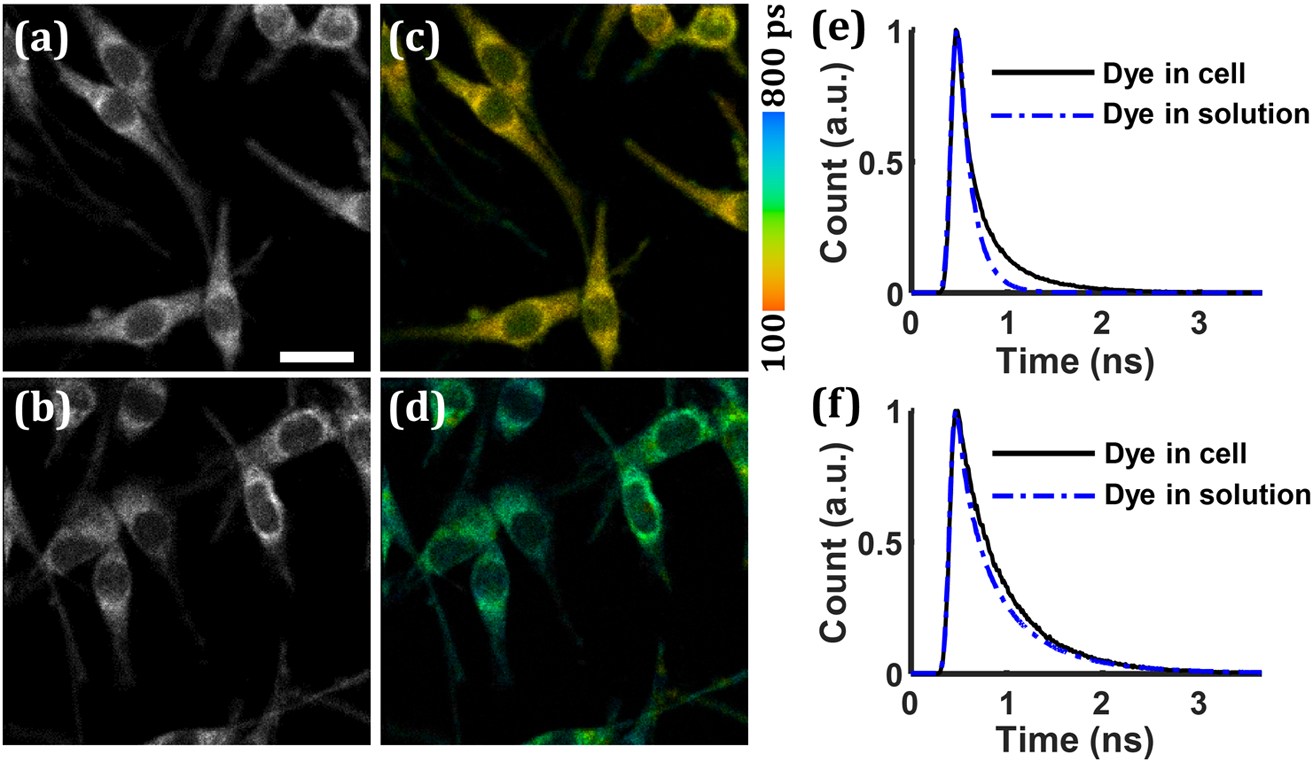}
    \caption{NIR-II confocal fluorescence lifetime imaging demonstrated using an SNSPD with a quantum efficiency centered at 1064~nm. Cultured C6 rat glioma cells were stained with (a,c) indocyanine green (ICG; fluorescence lifetime of 144.8 ps) or (b,d) human serum albumin-ICG. Panels (a) and (b) show NIR-II fluorescence intensity images, in which the two fluorophores are indistinguishable. Panels (c) and (d) display color-coded fluorescence lifetime images, allowing clear differentiation between the two fluorophores. Panels (e) and (f) present fluorescence decay curves of the fluorophores in cellular and solution environments. Scale bars in (a)–(d) are $20~\mu$m, figure adapted with permission from Yu et al  © Optica Publishing Group \cite{yu_intravital_2020}.}
    \label{fig:opticaFLIM}
\end{figure}

FLIM was demonstrated by Yu et al. \cite{yu_intravital_2020} in a confocal microscope equipped with an SNSPD using time-correlated single-photon counting, enabling intravital three-dimensional, multi-fluorophore imaging in the NIR-II region. Fluorescence lifetimes as short as 100~ps to be measured. As shown in Fig.~\ref{fig:opticaFLIM}, the lifetime of ICG or HSA-ICG (HSA: human serum albumin) could be clearly distinguished. 

In addition to FLIM, the sensitivity of SNSPDs has been demonstrated in time-resolved photoluminescence imaging \cite{buschmann_integration_2023}, with an SNR improvement by three orders of magnitude compared with a conventional PMT. Furthermore, singlet oxygen luminescence detection has been demonstrated with SNSPDs using optical-fiber delivery and collection \cite{gemmell_singlet_2013}, which is especially relevant for photodynamic therapy. Finally, the superior timing resolution of SNSPDs could further benefit TCSPC-based methods such as single-molecule detection \cite{barman_single-molecule_2025} and Förster resonance energy transfer (FRET) imaging \cite{ha_single-molecule_2001}.

\subsection{Quantum-Based Approaches to Microscopy}
Classical microscopy methods commonly rely on PMTs and SPADs for intensity measurements; however, these detectors also exhibit excellent performance in the photon-counting regime. Beyond the classical imaging domain, we can leverage concepts and principles from quantum physics to surpass classical limits. These approaches include the use of squeezed light \cite{taylor_biological_2013}, correlated photons \cite{tsao_enhancing_2025}, or entangled photons \cite{defienne_polarization_2021}. Such light sources enable label-free imaging and exploit higher-order correlation functions to surpass both the classical shot-noise limit and the classical resolution (diffraction) limit \cite{tenne_super-resolution_2019, schwartz_superresolution_2013}.

In addition, the emerging range of photon-counting detectors designed for analyzing quantum light sources can enhance imaging performance in itself. SNSPDs are becoming increasingly popular for quantum information processing, including quantum key distribution and optical computation \cite{you_superconducting_2020}. This is particularly pertinent for the life sciences, where photon budgets are inherently limited. Examples include imaging autofluorescence, deep-tissue imaging, and imaging in highly turbid media \cite{peterkovic_optimizing_2025} as already mentioned. In microscopy, SNSPDs can add value to systems exploring quantum-enhanced and correlation-based microscopy, where imaging contrast or spatial resolution arises from properties associated with the quantum nature of light \cite{zhang_quantum_2024}. 

\section{Integration Challenges and Solutions for SNSPD-Based Imaging Systems}
\label{sect:challenges}
The successful integration of any detector into imaging systems requires addressing several challenges that directly impact usability and clinical translation. In this perspective, we focus specifically on SNSPDs. Notably, state-of-the-art performance in one metric, as discussed in Section~\ref{sect:fundamental}, is often achieved by compromising others, reflecting unavoidable trade-offs \cite{venza_research_2025}. Below, we summarize the main barriers to biophotonics applications and highlight ongoing developments aimed at addressing these issues. 

\subsection{Optical Coupling Efficiency and Detector Active Area}
Optical coupling efficiency is particularly critical in biophotonics applications, where signals are inherently weak. Losses in optical coupling are a dominant source of inefficiency in SNSPD systems and are strongly dependent on the optical delivery method. In imaging applications, these involve the collection of light (e.g., fluorescence) and transmission to the SNSPD using both single-mode fibers (SMFs) and multimode fibers (MMFs), as well as free-space approaches. For SMF-coupled SNSPDs, efficiencies above 90$\%$ have been achieved using self-aligned packaging \cite{miller_compact_2011} or cryogenic nanopositioners \cite{marsili_detecting_2013}. MMF-coupled configurations can achieve $80$-$85\%$ system detection efficiency \cite{chang_multimode-fiber-coupled_2019} at a wavelength of 516~nm. By contrast, free-space coupling can support higher numerical apertures but is more sensitive to alignment and suffers from losses at the cryostat window and other optical interfaces. Reported efficiencies are typically around $60\%$ \cite{bellei_free-space-coupled_2016}. However, these parameters are highly dependent on the fibers used and on SNSPD structural parameters such as fill factor and active area. For more detail on how fiber type, fill factor, and active area jointly influence efficiency, the reader is referred to the following reviews \cite{dauler_review_2014, esmaeil_zadeh_single-photon_2017}. 

One approach to mitigating coupling loss is to increase the active area of single-pixel SNSPDs. Since the fiber core diameter sets a lower bound on the optical spot size, the detector must be larger than the fiber mode to achieve high coupling efficiency \cite{bellei_free-space-coupled_2016}. However, larger devices exhibit increased kinetic inductance \cite{kerman_kinetic-inductance-limited_2006}, which increases the detector recovery time and limits count rates \cite{rosenberg_high-speed_2013}. Fabrication of circular SNSPDs with diameters of 50~$\mu$m, coupled to multimode optical fibers, has yielded SDEs exceeding 80\% at 0.85~$\mu$m and 1.55~$\mu$m \cite{chang_multimode-fiber-coupled_2019}. Balancing active-area size with recovery dynamics remains an open optimization problem. 

For high-NA epi-detection, which is common in biophotonics, addressing these losses will be essential to ensure robust, high-fidelity signal collection.

\subsection{Large Array SNSPDs}
Expanding the active area by pixelating single-pixel SNSPDs is an alternative or complementary method for reducing coupling loss. Furthermore, the limited availability of pixelated SNSPDs is one reason why SNSPDs have been predominantly applied to point-scanning and have seen limited adoption in widefield imaging modalities. Smaller SNSPD arrays have already demonstrated advantages, as discussed in Section \ref{sect:opportunities}, particularly in confocal microscopy. Recently, Oripov et al. reported an SNSPD camera with 400,000 pixels \cite{oripov_superconducting_2023}, which could enable a widefield epifluorescence imaging platform. Large SNSPD arrays necessitate integrating cryogenic readout circuits close to the detectors and require physical separation from processing units that operate at higher temperatures. 

\subsection{Operating Temperature}
SNSPDs require operating temperatures well below their transition temperature, necessitating the use of a cryostat with precise thermal anchoring and vibration isolation. This requirement significantly increases the system's size, weight, power consumption, and cost (SWaP-C). The integration of SNSPDs into clinical settings has been limited by their SWaP-C \cite{robinson_portable_2023} and by noise generated by the operation of the cryostat and vacuum pump \cite{parfentyeva_fast_2023}. 

One method of reducing the SWaP-C of SNSPD systems is the development of miniaturized cryocoolers \cite{gemmell_miniaturized_2017, you_superconducting_2018, cao_progress_2020}. Such systems typically use a $^4$He Joule-Thomson (JT) stage. However, the cost of these cryocoolers is much higher than commercially adopted GM cryocoolers \cite{you_miniaturizing_2018}.

Alternatively, realizing next-generation high-temperature SNSPDs could substantially increase the allowable operating temperature of these devices \cite{charaev_single-photon_2024, merino_two-dimensional_2023}.
The operating temperature of the best-performing SNSPDs lies between 0.1~K and 2.8~K. However, single-photon detection has been demonstrated at 20~K \cite{charaev_single-photon_2024}, which would enable cryostats with at least half the SWaP-C compared to that of a 1~K system \cite{venza_research_2025}.

These developments are beginning to address the main limitations that have historically constrained the use of SNSPDs in biomedical imaging. As cryogenic and system-level barriers continue to be addressed, we anticipate that the path to clinically deployable SNSPD-based imaging systems will become increasingly viable.

\subsection{Polarization Sensitivity}
Several biophotonics modalities exploit polarization, including polarization reflectance spectroscopy, polarization imaging, and polarization-sensitive OCT \cite{tuchin_polarized_2016}. Standard meander-type nanowire geometries exhibit strongly polarization-dependent detection efficiency. This arises because the meander acts as a subwavelength grating \cite{natarajan_superconducting_2012}. Efficiency is maximized when the incident polarization is aligned with the nanowire and may decrease by up to 90\% for orthogonal polarization states \cite{sun_polarization_2021}. This dependence is particularly detrimental in fluorescence microscopy, where emission is depolarized by scattering and molecular dynamics \cite{liput_guide_2020}.
Mitigation strategies include dual-polarization collection using polarizing beam splitters \cite{xia_short-wave_2021} and novel device geometries such as spiral or fractal nanowires, which can reduce polarization dependence to near-unity levels \cite{meng_fractal_2022}. Development of such strategies can simplify integration into existing setups, for example, by enabling the use of a single SNSPD rather than dual-channel collection geometries. While these solutions are promising, widespread adoption in imaging platforms remains limited.
Conversely, SNSPDs with high polarization sensitivity have been applied to single-photon-level polarization discrimination \cite{guo_single_2015}, and may enable new contrast mechanisms by highlighting structural anisotropy or birefringence in biological samples \cite{boer_polarization_2017}.

\subsection{Dynamic Range}
The dynamic range of a detector limits the ability to measure both strong and weak signals simultaneously. SNSPDs suffer from a limited dynamic range, which constrains their performance in applications with large signal intensity variations. A prime example of such an application is OCT, which relies on the collection of backscattered light from structures with a contrasting refractive index to the surrounding material. A high dynamic range detection enables the collection of specular reflection from the surface of the sample, while maintaining efficient collection from deeper subsurface layers \cite{aumann_optical_2019}. The dynamic range is limited by the detector's MCR and refresh rate. Circuit-level advances in SNSPDs have increased the MCR to several gigahertz, with Tan et al. \cite{tan_achieving_2025} reporting a MCR of 3~GHz for a single-pixel SNSPD. Additionally, pixelated SNSPDs have improved the dynamic range by reducing the inductance, thereby increasing the MCR of the device to as high as 1.5~GHz \cite{resta_gigahertz_2023}. 
As discussed in Section~\ref{sect:fundamental}, an ideal SNSPD would continuously detect photons without disturbance. In practice, however, high count rates can trigger latching, in which the SNSPD remains trapped in a resistive state. As mentioned previously, electronic noise, thermal fluctuations, and environmental factors can contribute to the onset of this behavior. For example, stray light can induce latching even when the detected count rate is well below the MCR \cite{liu_nonlatching_2012}. Such instabilities reduce the applicability of SNSPDs in clinical translation. 

Count-rate saturation occurs when the detector reaches its MCR and begins to lose sensitivity to additional incoming photons due to limited recovery time between events. Both latching and saturation constrain the use of SNSPDs in applications requiring high dynamic range \cite{zheng_high_2025}, underscoring the need for quantitative characterization and mitigation strategies in high-flux applications.

These system-level characteristics remain active areas of development and are crucial for improving system robustness and facilitating clinical usability. Progress in optical and cryogenic technologies, alongside electronic integration, will be central to enabling the deployment of SNSPDs beyond research settings and into translational and clinical laboratories. 

\section{Outlook and Emerging Applications of SNSPDs in Biophotonics}
\subsection{Deep-Tissue Imaging in NIR-III and Beyond}
Imaging in the NIR region has traditionally been limited to wavelengths below 1.7~$\mu$m due to the lack of suitable fluorophores and high-efficiency detectors. The NIR-III window (2.08-2.34~$\mu$m), however, is predicted to offer the most favorable imaging performance across the NIR spectrum \cite{feng_perfecting_2021}. It has two distinct advantages: reduced photon scattering and increased absorption that can enhance image contrast \cite{feng_perfecting_2021}. When combined with novel fluorophores and SNSPD-based detection, the NIR-III range may enable unprecedented penetration depths and SNRs that were previously unattainable.

Beyond tissue-level studies, SNSPDs also hold promise for molecular-scale biophotonics in this spectral region. The capabilities of SNSPDs could extend advanced molecular contrast mechanisms, such as single-molecule Förster resonance energy transfer (FRET) \cite{ha_single-molecule_2001, liput_guide_2020} and localization-based super-resolution microscopy \cite{baddeley_biological_2018}, into spectral windows that penetrate far deeper into biological tissue. Together, these developments position SNSPD-enabled NIR-III imaging as a powerful avenue for biophotonics, scaling from whole-organ to single-molecule scale. 

\subsection{Gentle, Low-Photodamage Imaging}
Increasing illumination power can improve image quality, but the specimen's tolerance to light ultimately constrains its upper limit. Excess light can generate toxic free radicals, disrupt cellular function, and lead to irreparable specimen damage or cell death \cite{magidson_chapter_2013}. This phototoxic limit defines an upper bound on the achievable SNR without perturbing normal physiology. The high detection efficiency of SNSPDs enables equivalent image quality at lower excitation intensities, thereby reducing photodose and allowing gentle, long-term \textit{in vivo} time-lapse experiments. 
A recent demonstration of quantum-inspired OCT using SNSPDs achieved imaging at an optical power level of 10~pW \cite{kolenderska_quantum-inspired_2020}. Such light-efficient approaches highlight the potential of SNSPDs to extend high-resolution imaging into regimes compatible with ultralow-dose, minimally invasive biomedical applications. 

\section{Conclusion} % 0.5 page 
\label{sect:conc}
Photon-counting devices are becoming increasingly important in biophotonics, reflecting an opportunity to use sophisticated detectors to capture scarce photons. Within this context, photon-counting detectors provide access to biological information that would otherwise be unobtainable. Each detector carries its own advantages and disadvantages in both biophotonics and clinical settings, and the reader should consider which detector best matches the demands of their specific application. 

SNSPDs provide a unique combination of near-unity detection efficiency, picosecond timing resolution, and broadband spectral sensitivity that directly addresses the limitations of conventional detectors. These capabilities enable modalities that benefit from both extreme sensitivity and temporal precision, ranging from deep-tissue fluorescence microscopy to photon-starved biophotonics. Furthermore, SNSPDs have shown advantages in terms of improved signal-to-noise ratio and image fidelity, allowing imaging capabilities not possible with conventional detectors. Their broadband response additionally opens pathways for deeper tissue imaging through access to extended biological windows, including the NIR-III window. Moreover, these detectors' high sensitivity has the potential to enable low photodose imaging, supporting more capable, yet non-invasive diagnostic tools.

Despite remaining challenges in cryogenic integration, array scalability, and SWaP-C, rapid progress in addressing these limitations shows a promising path forward for translational imaging platforms. Over the coming decade, their adoption is expected to expand further from preclinical research to clinical imaging and spectroscopy. 

Ultimately, SNSPDs are poised to both enhance existing techniques and to unlock entirely new regimes of ultralow-dose high-resolution biomedical imaging. Their adoption may well define the next era of minimally invasive imaging. 

% \disclosures 
\subsection*{Disclosures}
The authors declare that there are no financial interests, commercial affiliations, or other potential conflicts of interest that could have influenced the objectivity of this research or the writing of this paper.

\subsection* {Code, Data, and Materials Availability} 
Data for Figure \ref{fig:OpticalSchematic} is publicly available in a Figshare repository: \url{https://adelaide.figshare.com/articles/figure/Supporting_information_for_Figure_3/30702932}.

\subsection* {Acknowledgments}
We acknowledge Adelaide Microscopy at the University of Adelaide, a Microscopy Australia (ROR: 042mm0k03) facility enabled by NCRIS. We thank Ben Sparkes and Ori Henderson-Sapir for the loan of equipment. Furthermore, we acknowledge the assistance and support received from Hein Zijlstra and Philipp Zolotov from Single Quantum. We thank Alexander Trowbridge for generating the 3D optical figure \ref{fig:OpticalSchematic}(a), and Ralf Mouthaan and Admir Bajraktarevic for their assistance in the laboratory. This work was supported by an Australian Research Council Laureate Fellowship (FL210100099).

%%%%% References %%%%%

\bibliography{report2}   % bibliography data in report.bib
\bibliographystyle{spiejour}   % makes bibtex use spiejour.bst

%%%%% Biographies of authors %%%%%

\vspace{2ex}\noindent\textbf{Emi Hughes} is a PhD candidate at the Centre of Light for Life at the University of Adelaide, Australia. She holds a Bachelor of Science (Honours) in computational physics. Her research interests include compressive fluorescence microscopy, structured-light microscopy, single-pixel imaging, confocal microscopy for deep-tissue imaging, and optical coherence tomography. 
\vspace{1ex}

\vspace{2ex}\noindent\textbf{Avinash Upadhya} is post-doctoral researcher working across both the Centre of Light for Life as well as the Reproductive Success Group at the Robinson Research Institute, University of Adelaide. He focuses on developing gentle optical imaging techniques for long-term monitoring of various biological samples, from single cells, to oocytes and embryos. For this, he employs advanced methods such as Bessel beam two-photon light sheet microscopy, hyperspectral imaging, and innovative acquisition strategies like compressed sensing. 

\vspace{1ex}

\vspace{2ex}\noindent\textbf{Kishan Dholakia} is Professor and Director of Centre of Light for Life at the University of Adelaide and Professor at the University of St Andrews, Scotland. His team uses structured (shaped) light fields especially for imaging, manipulation and precision measurement. He is a Fellow of the Royal Society of Edinburgh, Optica, SPIE and the Institute of Physics (IOP, UK). He has won a number of awards including the R.W. Wood Prize of Optica (2016), the IOP Thomas Young Medal and Prize (2017), the SPIE Dennis Gabor Award (2018) and the Australian Research Council Laureate Fellowship (2021). He is South Australian Scientist of the Year 2025.

\vspace{1ex}

\listoffigures
\listoftables
\end{document}